\documentclass[11pt]{article}
\usepackage{amsmath,amssymb,amsfonts}
\usepackage[english]{babel}
\makeatletter
\@addtoreset{equation}{section}
\makeatother

\def\nn{\nonumber} \def\bd{\begin{document}} \def\ed{\end{document}}
\def\ds{\documentstyle}
\let\bm=\bibitem
\newcommand{\be}{\begin{equation}}
\newcommand{\ee}{\end{equation}}
\newcommand{\bea}{\setlength\arraycolsep{2pt} \begin{eqnarray}}
\newcommand{\eea}{\end{eqnarray}}
\newcommand{\hoch}[1]{$\, ^{#1}$}
\def\p{\partial}

\title{\large {\bf Conserved charges of black holes in Weyl and Einstein-Gauss-Bonnet gravities}}
\date{}

\author{Jun-Jin Peng\footnote{pengjjph@163.com}  \\ \\
\small \sl Institute of Technical Physics, SEEE, Wuhan Textile University,\\
\small Wuhan, Hubei 430073, People's Republic of China \\
\small \sl Kavli Institute for Theoretical Physics China, \\
\small \sl Institute of Theoretical Physics, Chinese Academy of Sciences, \\
\small P.O.Box 2735, Beijing 100080, People's Republic of China
}

\begin{document}

\maketitle
\vspace{20pt}

\begin{center}
\textbf{Abstract}
\end{center}
An off-shell generalization of the Abbott-Deser-Tekin (ADT) conserved charge was recently
proposed by Kim et al. They achieved this by introducing off-shell Noether currents and potentials.
In this paper, we construct the crucial off-shell Noether current by the variation of the Bianchi
identity for the expression of motion equation, with the help of the property of Killing vector.
Our Noether current, which contains an additional term that is just one half of the Lie derivative
of a surface term with respect to the Killing vector, takes a different form in comparison with the
one in their work. Then we employ the generalized formulation to calculate the
quasi-local conserved charges for the most general charged spherically symmetric and the dyonic
rotating black holes with AdS asymptotics in four-dimensional conformal Weyl gravity,
as well as the charged spherically symmetric black holes in arbitrary dimensional
Einstein-Gauss-Bonnet gravity coupled to Maxwell or nonlinear electrodynamics in AdS spacetime.
Our results confirm those through other methods in the literature.

\voffset=-.90pt
\vspace{40pt}

\section{Introduction}

Modified gravity theories that involves higher curvature terms in the Lagrangian have been
extensively investigated, generally motivated by the intriguing feature that these
higher curvature terms render the gravity theories perturbatively renormalizable in the
quantization process \cite{RenormStell}. A very natural higher-order derivative modification of
general relativity is the fourth-order theories of gravitation, which includes the well-known
theories of Weyl gravity and Einstein-Gauss-Bonnet gravity. To the former, its Lagrangian
contains the square of the Weyl tensor, so it is invariant under the local conformal
transformation of the metric. The Lagrangian for Einstein-Gauss-Bonnet gravity includes up to
the term with quadratic Riemann tensor, which can be thought as the higher curvature correction
to general relativity in the low energy limit of heterotic string theory. Due to the salient
properties of both the gravity theories, together with the AdS/CFT correspondence, a lot of
efforts have been devoted to seeking asymptotically AdS black hole solutions in Weyl gravity
\cite{StaticRWeyl,LLuCRotSW} and Einstein-Gauss-Bonnet gravity \cite{GaussBsoBD,GaussBsoWh,
GaussBsoWi,GaussBsoCai,GaussBsoCNO}, to provide various interesting backgrounds of spacetime.
Generally speaking, after obtaining a black hole solution, an important task is to identify
its conserved charges, such as the energy and the angular momentum.

Till now several approaches have been proposed to compute the conserved charges of asymptotically
AdS solutions, such as the so-called counterterm subtraction approach \cite{BrownYBK} generalized
from Brown-York method \cite{BroYMass}, the Ashtekar-Magnon-Das formalism \cite{AMDmass}, the
covariant phase space approach \cite{IyerWald}, the method \cite{BarnichB,Barnich,BarnichC}
developed by Barnich et al. and the Abbott-Deser-Tekin (ADT) formalism \cite{AbbottD,DeserT}.
Particularly, the ADT formalism, which is defined by the Noether potential got through the
linearized perturbation for the expression of motion equation in a fixed background of AdS
spacetime, has made some progress on computation scheme for conserved charges of asymptotically
AdS black holes in fourth-order gravity theories. Since the background metric is a vacuum
solution of equation of motion (EOM), the Noether potential in ADT formalism is on-shell.
Recently, in Ref. \cite{KimKY}, Kim, Kulkarni and Yi proposed a quasi-local formulation of
conserved charges by generalizing the on-shell Noether potential in the ADT formalism to
off-shell level, as well as following works \cite{BarnichB,Barnich,BarnichC} to incorporate a single
parameter path in the space of solutions into their definition. These modifications
make it more operable to evaluate the Noether potential in terms of the corresponding current.
The generalized formalism for the quasi-local conserved charges provides a more efficient way
to compute the ADT conserved charges for covariant theories of gravity, and it has been extended
to the theory of gravity with a gravitational Chern-Simons term \cite{CiteChernS} and the gravity
theory in the presence of matter fields \cite{CiteHJPY}. In \cite{CiteLifBH}, it was utilized to
obtain the mass of the three- and five-dimensional Lifshitz black holes. To compare with the
original ADT formalism, it is meaningful to employ this generalized quasi-local formulation to
study the conserved charges in higher-order derivative gravity theories.

In this paper, to provide a deep understanding on the generalized ADT formalism proposed in
\cite{KimKY}, we derive the off-shell Noether current that educes the Noether potential
finally entering into the formulation of conserved charges from different perspective. Our
derivation endows the off-shell Noether current with a natural connection with its corresponding
potential. Then we extend this formalism to investigate the quasi-local conserved charges of
charged (rotating) black holes with AdS asymptotics in the two typical fourth-order derivative
gravity theories: conformal Weyl gravity and Einstein-Gauss-Bonnet gravity. The remainder of
this paper goes as follows. In section \ref{secone}, we give a brief review of the method in
\cite{KimKY}. However, unlike there, we derive the off-shell Noether current and its corresponding
potential through the variation of the Bianchi identity for the expression of motion equation.
Our results are formally different from those in \cite{KimKY}. In section \ref{sectwo}, we first
present the explicit expressions of the off-shell Noether potentials in Weyl gravity. Then these
quantities are applied to compute the mass of the most general static black hole and both the mass
and angular momentum of the dyonic rotating black hole in four-dimensional Weyl gravity. In
section \ref{secthree}, we calculate the energy of the general charged spherically symmetric black
hole in arbitrary dimensional Einstein-Gauss-Bonnet gravity, coupled to Maxwell or nonlinear
electrodynamics in AdS spacetime. The general formalizations of the Noether potentials
for Einstein-Gauss-Bonnet gravity are also given. The last section is our conclusions.

\section{The generalized ADT formalism}\label{secone}
In this section, we shall review the formulation of conserved charges in \cite{KimKY},
which can be thought as the off-shell extension of the ADT formalism since both the
Noether current and potential there were constructed without the requirement
that the gravitational fields must satisfy the motion equation. What is more, another
obvious difference from the ADT formalism is that one parameter path in the solution
space was introduced to present the final definition of the quasi-local conserved
charge in \cite{KimKY}. However, unlike in \cite{KimKY}, where the start point to
derive the off-shell Noether current and potential is the variation and diffeomorphism
transformation of the action, we shall give a different derivation of the two quantities
by directly varying the Bianchi identity and using the property of Killing vector.
We proceed by considering the Lagrangian of a $D$-dimensional generally diffeomorphism
covariant gravity theory \cite{IyerWald}
\be
\mathcal{L}= \sqrt{-g}L[g_{\mu\nu},R,R_{\mu\nu},R_{\mu\nu\rho\sigma},
\nabla R,\nabla R_{\mu\nu},\nabla R_{\mu\nu\rho\sigma},\cdot\cdot\cdot]
\, , \label{GenLagran}
\ee
which includes no other matter fields. The variation of Eq. (\ref{GenLagran}) generally
yields
\be
\delta\mathcal{L}=\sqrt{-g}\mathcal{E}_{\mu\nu}\delta g^{\mu\nu}+\sqrt{-g}
\nabla_\mu \Theta^\mu(g;\delta g) \, , \label{VarGenLagran}
\ee
where $\mathcal{E}_{\mu\nu}$ is the expression of equation of motion (EOM), and
$\Theta^\mu(g;\delta g)$
\footnote{It multiplied by $\sqrt{-g}$ is equal to the surface term $\Theta^\mu(g;\delta g)$
defined in \cite{KimKY}.}
is a surface term. To preserve diffeomorphism, $\mathcal{E}_{\mu\nu}$ satisfies
the Bianchi identity
\be
\nabla_\mu \mathcal{E}^{\mu\nu} =0 \label{BianchI}\, .
\ee

Varying the Bianchi identity (\ref{BianchI}), we get
\be
\nabla_\mu\delta\mathcal{E}^{\mu\nu}
+\delta\Gamma^\mu_{\mu\lambda}\mathcal{E}^{\lambda\nu}
+\delta\Gamma^\nu_{\mu\lambda}\mathcal{E}^{\mu\lambda}
=0 \, .
\ee
multiplying the above equation by a Killing vector $\xi^\mu$, we further obtain
\be
\nabla_\mu\Big(\delta\mathcal{E}^{\mu\nu}\xi_\nu
+\mathcal{E}^{\mu\nu}\xi^\lambda\delta g_{\nu\lambda}
+\frac{1}{2}\mathcal{E}^{\mu\nu}\xi_\nu g^{\alpha\beta}\delta g_{\alpha\beta}
-\frac{1}{2}\xi^\mu\mathcal{E}^{\alpha\beta}\delta g_{\alpha\beta} \Big)
+\frac{1}{2}\delta g_{\alpha\beta}\mathcal{L}_\xi\mathcal{E}^{\alpha\beta}
=0 \, , \label{VarBIone}
\ee
where $\mathcal{L}_\xi$ denotes the Lie derivative with respect to the Killing vector
$\xi^\mu$. In order to get Eq. (\ref{VarBIone}), the equation of Killing vector
$\nabla_{(\mu}\xi_{\nu)}=0$ is used and the variation of the gravitational field
$g_{\mu\nu}\rightarrow g_{\mu\nu}+\delta g_{\mu\nu}$ is assumed to
preserve the Killing vector, namely, $\delta\xi^\mu=0$, here and in the following.
Obviously, the component in the bracket of Eq. (\ref{VarBIone}) is just the off-shell
Noether current presented in \cite{KimKY,BClement}. This makes it possible to construct
the Noether current from Eq. (\ref{VarBIone}). Actually, combined with the equation
\be
\mathcal{L}_\xi\delta(\sqrt{-g}L)
=-\sqrt{-g}\delta g_{\mu\nu}\mathcal{L}_\xi\mathcal{E}^{\mu\nu}+\sqrt{-g}
\nabla_\mu [\mathcal{L}_\xi\Theta^\mu(g;\delta g)]
=0
\, ,
\ee
which is the Lie derivative of Eq. (\ref{VarGenLagran}) in terms of the Killing
vector $\xi^\mu$, one can rewrite Eq. (\ref{VarBIone}) as
\be
\nabla_\mu\Big[\delta\mathcal{E}^{\mu\nu}\xi_\nu
+\mathcal{E}^{\mu\nu}\xi^\lambda\delta g_{\nu\lambda}
+\frac{1}{2}\mathcal{E}^{\mu\nu}\xi_\nu g^{\alpha\beta}\delta g_{\alpha\beta}
-\frac{1}{2}\xi^\mu\mathcal{E}^{\alpha\beta}\delta g_{\alpha\beta}
+\frac{1}{2}\mathcal{L}_\xi\Theta^\mu(g;\delta g)\Big]=0 \, . \label{CovofshNoeCur}
\ee
Thus, the off-shell Noether current can be defined by
\be
\mathcal{J}^\mu=\delta\mathcal{E}^{\mu\nu}\xi_\nu
+\mathcal{E}^{\mu\nu}\xi^\lambda\delta g_{\nu\lambda}
+\frac{1}{2}\mathcal{E}^{\mu\nu}\xi_\nu g^{\alpha\beta}\delta g_{\alpha\beta}
-\frac{1}{2}\xi^\mu\mathcal{E}^{\alpha\beta}\delta g_{\alpha\beta}
+\frac{1}{2}\mathcal{L}_\xi\Theta^\mu(g;\delta g)
 \,. \label{OffShNouC2}
\ee
Note that the last term in the above equation, which is one half of the Lie derivative
of the surface term $\Theta^\mu(g;\delta g)$ with respect to the Killing vecotr $\xi^\mu$,
is an additional term compared with the Noether current in \cite{KimKY,BClement}. Of course,
it is feasible to define the off-shell Noether current $\mathcal{J}^\mu$ as same as the one
there in terms of Eq. (\ref{CovofshNoeCur}). One only needs to add a term
$-\nabla_\mu \big[\delta(\sqrt{-g}\Theta^\mu(g;\xi))/\sqrt{-g}\big]/2$ to the left side of
Eq. (\ref{CovofshNoeCur}), together with the result
$\mathcal{L}_\xi(\sqrt{-g}\Theta^\mu(g;\delta g))-\delta(\sqrt{-g}\Theta^\mu(g;\xi))=0$
given in \cite{IyerWald}. To see this clearly, under the diffeomorphism $\xi^\mu$, the metric
$g_{\mu\nu}$ transforms as $\delta g_{\mu\nu}=\mathcal{L}_\xi g_{\mu\nu} =0$ and
$\delta(\sqrt{-g}L)=\mathcal{L}_\xi(\sqrt{-g}L)=0$, so Eq. (\ref{VarGenLagran}) becomes
$\nabla_\mu \Theta^\mu(g;\xi) =0$, whose variation leads to
\be
\delta(\nabla_\mu \Theta^\mu(g;\xi))
=\nabla_\mu \Big[\frac{1}{\sqrt{-g}}\delta(\sqrt{-g}\Theta^\mu(g;\xi))\Big] =0 \, . \nn
\ee

Next, we shall derive the off-shell Noether potential $Q_{ADT}^{\mu\nu}$, which is
associated with the off-shell Noether current $\mathcal{J}^\mu$ through the relation
\be
\mathcal{J}^\mu=\nabla_\nu Q_{ADT}^{\mu\nu}
\, . \label{GenOfShNP}
\ee
To do this, substituting the Lie derivative of the surface term $\Theta^\mu(g;\delta g)$
\be
\mathcal{L}_\xi\Theta^\mu(g;\delta g)= -2\nabla_\nu
\xi^{[\mu}\Theta^{\nu]}(g;\delta g)
+\xi^\mu\nabla_\nu\Theta^\nu(g;\delta g) \,
\ee
and Eq. (\ref{VarGenLagran}) into the off-shell current (\ref{OffShNouC2}), we present
the current $\mathcal{J}^\mu$ as the form
\be
\sqrt{-g}\mathcal{J}^\mu
=\delta\Big(\sqrt{-g}\mathcal{E}^{\mu\nu}\xi_\nu
+\frac{1}{2}\sqrt{-g}\xi^\mu L\Big)
-\sqrt{-g}\nabla_\nu\xi^{[\mu}\Theta^{\nu]}(g;\delta g) \, .
\ee
By following \cite{TPadman} to define another off-shell current $J^\mu$ as
\be
J^\mu=2\mathcal{E}^{\mu\nu}\xi_\nu+\xi^\mu L
=\nabla_\nu K^{\mu\nu}
\, , \label{AnXiNoethC}
\ee
where $K^{\mu\nu}$ is the off-shell Noether potential corresponding to
the current $J^\mu$ and it is easy to verify that $\nabla_\mu J^\mu=0$, we further
send the current $\mathcal{J}^\mu$ into the form
\be
\sqrt{-g}\mathcal{J}^\mu
=\partial_\nu\Big[\frac{1}{2}\delta(\sqrt{-g} K^{\mu\nu})
-\sqrt{-g}\xi^{[\mu}\Theta^{\nu]}(g;\delta g)\Big]
=\partial_\nu\big(\sqrt{-g} Q_{ADT}^{\mu\nu} \big)
\, .\ee
The above equation yields the important relation between ADT potential
and the off-shell Noether potential $K^{\mu\nu}$, namely,
\be
\sqrt{-g}Q_{ADT}^{\mu\nu}
=\frac{1}{2}\delta(\sqrt{-g}K^{\mu\nu})
-\sqrt{-g}\xi^{[\mu}\Theta^{\nu]}(g;\delta g) \, , \label{GenOffShNP1}
\ee
or equally,
\be
Q_{ADT}^{\mu\nu}
=\frac{1}{2}\delta K^{\mu\nu}
+\frac{1}{4}K^{\mu\nu}g^{\alpha\beta}\delta g_{\alpha\beta}
-\xi^{[\mu}\Theta^{\nu]}(g;\delta g) \, . \label{GenOffShNP2}
\ee

In comparison with the results in \cite{KimKY}, although the off-shell Noether currents
$\mathcal{J}^\mu$ and $J^\mu$ obtained by the variation of Bianchi identity (\ref{BianchI})
are formally different from their corresponding quantities there,  each of them is
correspondingly equal since one can verify that $\Theta^{\mu}(g;\xi)=\mathcal{L}_\xi\Theta^\mu(g;\delta g)
=\mathcal{L}_\xi\mathcal{E}^{\mu\nu}=0$ for the generally diffeomorphism covariant
Lagrangian (\ref{GenLagran}). The merit of our formulation for the Noether current
$\mathcal{J}^\mu$ is that it becomes more natural to get the off-shell Noether potential
$Q_{ADT}^{\mu\nu}$. In fact, to finally get the relation between $Q_{ADT}^{\mu\nu}$
and $K^{\mu\nu}$, the condition
$\mathcal{L}_\xi(\sqrt{-g}\Theta^\mu(g;\delta g))-\delta(\sqrt{-g}\Theta^\mu(g;\xi))=0$
related to the symplectic current \cite{IyerWald} has
been used in \cite{KimKY}, but it is not needed in our case. Particularly, when the
background metric satisfies the vacuum motion equation: $\mathcal{E}_{\mu\nu}=0$,
all the Noether currents and potentials become the conventional ones in ADT formalism
\cite{AbbottD,DeserT}.

Finally, one can propose a formulation of the conserved charge in terms of the integral of
the Noether potential (\ref{GenOffShNP2}) over the boundary of a spatial hypersurface under
the conditions that a background metric that is a vacuum solution of EOM is fixed and the
perturbation of the given metric is taken as the divergence between it and the fixed background
metric,  like the ADT method \cite{AbbottD,DeserT}. However, Ref. \cite{KimKY} gave
another definition by following works \cite{BarnichB,Barnich,BarnichC} to incorporate
a single parameter path characterized by a parameter $s$ $(0\leq s\leq 1)$ in the space
of solutions. This path interpolates between the given solution and the background solution
through parameterizing a set of free parameters $\mathcal{C}$ in the space for the solutions
of EOM as $s\mathcal{C}$. On basis of the Noether potential $Q_{ADT}^{\mu\nu}$ in
Eq. (\ref{GenOffShNP1}), by integrating the variable $s$, one can define the quasi-local
conserved charge by
\bea
\mathcal{Q}&=&\frac{1}{8\pi}\int_0^1 ds \int d\Sigma_{\mu\nu} Q_{ADT}^{\mu\nu}(g;s) \nn \\
&=&\frac{1}{16\pi}\int d^{(D-2)}x_{\mu\nu} \triangle \hat{K}^{\mu\nu}
-\frac{1}{8\pi}\int_0^1ds\int d\Sigma_{\mu\nu}\xi^{[\mu}\Theta^{\nu]}(g;s)
\, , \label{QdefineAn}
\eea
where $d\Sigma_{\mu\nu}=\sqrt{-g}d^{(D-2)}x_{\mu\nu}=\frac{1}{2}\frac{1}{(D-2)!}
\epsilon_{\mu\nu\mu_1\mu_2\cdot\cdot\cdot\mu_{(D-2)}}dx^{\mu_1}\wedge\cdot\cdot\cdot
\wedge dx^{\mu_{(D-2)}}$ and
\be
\triangle \hat{K}^{\mu\nu} =\sqrt{-g}K^{\mu\nu}\big|_{s=1}-\sqrt{-g}K^{\mu\nu}\big|_{s=0} \nn
\ee
is the finite difference between the given solution and the background solution, i.e.
the two end points of the single parameter path. Eq. (\ref{QdefineAn}) can be a proposal
of the conserved charge, defined in the interior region or at the asymptotical infinity,
for any covariant gravity theory with the Lagrangian (\ref{GenLagran}) whenever its
integration is well-defined. In the following sections, we shall make use of Eq. (\ref{QdefineAn})
to calculate the mass and angular momenta of charged static and rotating black holes in
Weyl and Einstein-Gauss-Bonnet gravities although Eq. (\ref{QdefineAn}) is defined
in terms of the Lagrangian for pure gravity, without any matter fields. We can do this since
the terms associated with the gauge fields fall off fast enough at asymptotic infinity to
guarantee that the integration is finite.

\section{Conserved charges of black holes in four-dimensional Weyl gravity}\label{sectwo}

In this section, we make use of the generalized ADT formalism in the previous section
to calculate the quasi-local conserved charges of the most general charged spherically
symmetric black hole and the charged rotating black hole in four-dimensional Weyl gravity.
The Lagrangian for Weyl gravity takes the form
\be
\mathcal{L}_W=\frac{1}{2}\alpha\sqrt{-g}C^2
=\frac{1}{2}\alpha \Big(R^{\mu\nu\rho\sigma}R_{\mu\nu\rho\sigma}-2R^{\mu\nu}R_{\mu\nu}
+\frac{1}{3}R^2\Big) \, , \label{LagraWeylG}
\ee
where $\alpha$ is a coupling constant, and the Weyl tensor $C_{\mu\nu\rho\sigma}$
is given by
\be
C_{\mu\nu\rho\sigma}=-(g_{\mu[\rho}R_{\sigma]\nu}-g_{\nu[\rho}R_{\sigma]\mu})
+\frac{1}{3}Rg_{\mu[\rho}g_{\sigma]\nu}+R_{\mu\nu\rho\sigma}
\ee
in four dimensions. The Weyl tensor has the same symmetry properties as Riemann curvature
but it is traceless, i.e. $C^\rho_{~\mu\rho\nu}=0$. The expression of EOM from the
Lagrangian  (\ref{LagraWeylG}) is
\be
\mathcal{E}^W_{\mu\nu}=2\alpha B_{\mu\nu}
=-\alpha(2\nabla^\rho\nabla^\sigma
+R^{\rho\sigma})C_{\mu\rho\sigma\nu}
\, ,
\ee
where $B_{\mu\nu}$ is just the Bach tensor. By using the properties for Lie derivative along
a Killing vector $\nabla_\mu \mathcal{L}_\xi =\mathcal{L}_\xi\nabla_\mu$ and
$\mathcal{L}_\xi R_{\mu\nu\rho\sigma}=0$, one can check that
$\mathcal{L}_\xi \mathcal{E}^W_{\mu\nu}=0$.

Now we present some quantities tightly related to our calculation, such as
the surface term $\Theta_W^{\mu}(g;\delta g)$ and the off-shell Noether
potentials $K_W^{\mu\nu}$, $Q_{W}^{\mu\nu}$ for the four-dimensional Weyl gravity.
More general results for higher derivative gravity theories can be found in
\cite{KimKY,CardosoWM}. The surface term $\Theta_W^{\mu}(g;\delta g)$ and the
potential $K_W^{\mu\nu}$ are read off as
\bea
\Theta_W^{\mu}(g;h)&=& 2\alpha C^{\mu\nu\rho\sigma}\nabla_\sigma h_{\nu\rho}
+2\alpha h_{\nu\sigma} \Big(\nabla^{[\mu}R^{\nu]\sigma}
+\frac{1}{6}g^{\sigma[\mu}\nabla^{\nu]}R\Big) \, , \label{WeylGsurf} \\
K_W^{\mu\nu}&=& 2\alpha C^{\mu\nu\rho\sigma}\nabla_{[\rho}\xi_{\sigma]}
-4\alpha \xi_\sigma \Big(\nabla^{[\mu}R^{\nu]\sigma}
+ \frac{1}{6}g^{\sigma[\mu}\nabla^{\nu]}R\Big) \, , \label{WeylGPK}
\eea
where and in what follows $h_{\mu\nu}\equiv\delta g_{\mu\nu}$ denotes the variation
of the metric and its indices are lowered or raised by the background metric
$g_{\mu\nu}$ or $g^{\mu\nu}$. It is easy to verify that $\Theta_W^{\mu}(g;\xi)=0$ when
$h_{\mu\nu}=\mathcal{L}_\xi g_{\mu\nu}=0$ and $\mathcal{L}_\xi \Theta_W^{\mu}(g;h)=0$.
The off-shell Noether potentials $Q_{W}^{\mu\nu}$ is
\be
Q_{W}^{\mu\nu}
=\frac{1}{2}\delta K_W^{\mu\nu}
+\frac{1}{4}hK_W^{\mu\nu}-\xi^{[\mu}\Theta_W^{\nu]}(g;h) \, , \label{WeylGNoeQ}
\ee
where $h$ in the second term denotes $h=g^{\alpha\beta}\delta g_{\alpha\beta}$.
For the explicit expression of $\delta K_W^{\mu\nu}$ see Eq. (\ref{VariWeylGK}) in the
appendix.

\subsection{The Conserved charge of the charged spherically symmetric black hole}\label{sebsetwoone}

In this subsection, we take into account of the quasi-local charge of the charged
black hole in static case. The total Lagrangian is
$\mathcal{L}_{total}=\mathcal{L}_W+L_{EM}$, where
$L_{EM}=\frac{1}{3}\alpha \sqrt{-g}F_{\mu\nu}F^{\mu\nu}$ is the Lagrangian for the
gauge field $A_\mu$ and the field strength $F_{\mu\nu}=2\partial_{[\mu}A_{\nu]}$.
The most general charged spherically symmetric black hole in this four-dimensional
Weyl gravity theory has the form \cite{StaticRWeyl}
\bea
ds^2&=&-f(r)dt^2+\frac{dr^2}{f(r)}+r^2(d\theta^2+\sin^2\theta d\phi^2) \, ,
\quad A=-\frac{q}{r}dt \, , \nn \\
f(r)&=& m+\frac{b}{r}+ar-\frac{1}{3}\Lambda r^2 \, , \label{SpheBHWeylG}
\eea
where the four constants $(a,b,m,q)$ is constrained by $3ab+1+q^2=m^2$.
When $a\neq 0$, which is the case we fist consider, we get $b=(m^2-1-q^2)/(3a)$
from the constraint.

To make use of the formulation (\ref{QdefineAn}) to calculate the conserved
charge of the static black hole (\ref{SpheBHWeylG}), we choose an infinitesimal
parametrization of a single parameter path in the solution space by letting the
constants $(m,q)$ change as
\be
m\rightarrow m+dm \, , \quad \quad q\rightarrow q+dq \, . \label{InPaSBHWeylG}
\ee
Under such a parametrization and the choice of the Killing vector $\xi^\mu=(-1,0,0,0)$,
the $(t,r)$ components of the Noether potentials
$K_W^{\mu\nu}$ and $Q_{W}^{\mu\nu}$ is given by
\bea
\sqrt{-g}K_W^{tr}&=&\frac{2\alpha
\sin\theta[2\Lambda (m^2-q^2-1)+3a^2(m-1)]}{9a}+\mathcal{O}(\frac{1}{r})
\, , \nn \\
\sqrt{-g}Q_W^{tr}&=&\frac{\alpha(3a^2+4m\Lambda)\sin\theta dm}{9a}
-\frac{4\alpha q\Lambda \sin\theta dq}{9a}
+\frac{2\alpha q\sin\theta dq}{3r} \, .
\eea
Then the mass of the charged spherically symmetric black hole (\ref{SpheBHWeylG})
can be computed as
\be
M=\frac{1}{4}\int \int_0^\pi \big(\lim_{r=\infty}\sqrt{-g}Q_W^{tr}\big) d\theta
=\frac{\alpha(3ma^2+2\Lambda m^2-2q^2\Lambda)}{18a} \, . \label{MasstaticBH}
\ee
The temperature $T$ and the entropy $S$ \cite{CLPstaticW} of the black hole are
\bea
T&=& \frac{1}{12\pi}\frac{1+q^2-m^2+3a^2r_+^2-2a\Lambda r_+^3}{ar_+^2} \, ,\nn \\
S&=&\frac{2\pi\alpha[1+q^2-m^2-(m+2)ar_+]}{3ar_+} \, , \nn
\eea
respectively, where $r_+$ is the event horizon, given by $f(r_+)=0$. Both the electric
charge $Q_e$ and the electric potential $\Phi_e$ are
\be
Q_e=-\frac{1}{3}\alpha q \, , \quad \Phi_e=\frac{q}{r_+} \, . \nn
\ee
One can verify that the mass (\ref{MasstaticBH}) satisfies the first law:
\be
dM=TdS+\Phi_e dQ_e \, .
\ee

The above first law shows that the generalized ADT formulation (\ref{QdefineAn}) is
applicable to the static black hole in four-dimensional Weyl gravity. However,
in \cite{LuPPconsW}, it was claimed that ADT formalism fails to give a finite result
if one trivially chooses the static AdS metric as the background to calculate the mass
of the black hole (\ref{SpheBHWeylG}) when $a\neq 0$, so the standard Noether method to
the Lagrangian of Weyl gravity was adapted, namely, they directly used the Noether potential
$K_W^{\mu\nu}$ to define the conserved charge as
\be
\hat{\mathcal{Q}}=\frac{1}{16\pi}\int_{r=\infty} d\Sigma_{\mu\nu} K_W^{\mu\nu}
\, . \label{LPPCCwyeG}
\ee
In fact, due to the disappearance of the $(t,r)$  component of the surface term
$\xi^{[\mu}\Theta_W^{\nu]}(g;h)$ at $r=\infty$, one can only utilize the Noether
potential $K_W^{\mu\nu}$, like Eq. (\ref{LPPCCwyeG}), to calculate the mass of
the static black hole, but there exists a divergence
$M_0=\hat{M}-M=(3a^2+2\Lambda)/(18a)$ between the mass $M$ and the one $\hat{M}$
got through the definition (\ref{LPPCCwyeG}), since the quantity
$\sqrt{-g}K_W^{tr}\big|_{m,q=0}$ does not vanish at $r=\infty$. In the uncharged
case, the mass $\hat{M}$ agrees with the one computed from the conserved current
that consists of the holographic response functions in \cite{GILovM}.

At the end of this subsection, we take into account the $a=0$ case. For convenience,
we recast the function $f(r)$ as
\be
f(r)= \sqrt{1+q^2}-\frac{2m}{r}-\frac{1}{3}\Lambda r^2 \, . \nn
\ee
Performing the parallel analysis as the case where $a\neq 0$, we obtain the mass
$M=-2\alpha m\Lambda/3$, which can also be got through the ADT formalism and the
Ashtekar-Magnon-Das method \cite{AMDmass,ConWPang}.

\subsection{Conserved charges of the dyonic rotating black holes}\label{subsectwotwo}

In four-dimensional conformal Weyl gravity with the Lagrangian
$\mathcal{L}_{total}=\mathcal{L}_W+L_{EM}$, the charged rotating
black hole solution, found in \cite{LLuCRotSW}, takes the form
\bea
ds^2&=& -\frac{\Delta}{\rho^2(r,\theta)}\Big(dt-\frac{a\sin^2\theta}{\Xi}d\phi\Big)^2
+\frac{\rho^2(r,\theta)}{\Delta}dr^2+\frac{\rho^2(r,\theta)}{F(\theta)}d\theta^2 \nn \\
&&+\frac{F(\theta)\sin^2\theta}{\rho^2(r,\theta)}\Big(adt-\frac{r^2+a^2}{\Xi}d\phi\Big)^2
\nn \\
\Delta &=& (r^2+a^2)(1+r^2\ell^2)-2mr+\frac{1}{6}m(p^2+q^2)r^3 \, ,\nn \\
\rho^2(r,\theta)&=& r^2+a^2\cos^2\theta \, , \quad
F(\theta)= 1-a^2\ell^2\cos^2\theta \, , \nn \\
\Xi &=& 1-a^2\ell^2 \, , \label{CRBHWeylG}
\eea
where the constants $a,m,p,q$ denotes the mass, angular momentum, magnetic and electric
charges, respectively. When $m=p=q=0$, the metric (\ref{CRBHWeylG}) becomes the conventional
$AdS_4$ spacetime, with the negative cosmological constant $\Lambda =-3\ell^2$.
Comparing this black hole solution with the conventional four-dimensional Kerr-Newman-AdS
solution in Einstein-Maxwell gravity, one finds that the last term related to the magnetic
and electric charge parameters $(p,q)$ in the function $\Delta$ is not the usual combination
$p^2+q^2$ for the Kerr-Newman-AdS black hole. Such a difference makes the solution
(\ref{CRBHWeylG}) has some new interesting properties \cite{CLPstaticW,HawkRDeng}.

We first calculate the energy and angular momentum of the black hole (\ref{CRBHWeylG})
in neutral case, namely, $p=q=0$. In such a case, we take an infinitesimal parametrization
of a single parameter path by letting the constants $(m,a)$ fluctuate as
\be
m\rightarrow m+dm \, , \quad \quad a\rightarrow a+da \, . \nn
\ee
In addition to the timelike Killing vector $\xi_{WM}^\mu=(-1,0,0,0)$, the $(t,r)$
components of the Noether potentials related to the energy are given by
\bea
\sqrt{-g}K_{WM}^{tr}&=&
\frac{4\alpha m\ell^2(3\sin^2\theta+3\Xi\cos^2\theta-\Xi)\sin\theta}{\Xi^2}
+\mathcal{O}(\frac{1}{r^2})
\, , \nn \\
\sqrt{-g}Q^{tr}_{WM}&=&
\frac{6\alpha am\ell^4[4-\Xi-(4-3\Xi)\cos^2\theta]\sin\theta da}{\Xi^3}\nn \\
&&+\frac{2\alpha\ell^2[3-\Xi-3(1-\Xi)\cos^2\theta]\sin\theta dm}{\Xi^2}
+\mathcal{O}(\frac{1}{r^2})
\, .
\eea
Utilizing the definition of the quasi-local conserved charge (\ref{QdefineAn}),
we obtain the energy of the neutral rotating black hole
\be
M_{NR}=\frac{2m\alpha\ell^2}{\Xi^2} \, .
\ee
To calculate the angular momentum, the spacelike Killing vector is chosen as
$\xi_{WJ}^\mu=(0,0,0,1)$. Then the $(t,r)$ components of the Noether potentials
related to the angular momentum $J_{NR}$ are presented by
\bea
\sqrt{-g}K_{WJ}^{tr}&=&\frac{12\alpha ma\ell^2\sin^3\theta}{\Xi^2}
+\mathcal{O}(\frac{1}{r^2})\, , \nn \\
\sqrt{-g}Q_{WJ}^{tr}&=&\frac{6\alpha m\ell^2(4-3\Xi)\sin^3\theta da}{\Xi^3}
+\frac{6\alpha a\ell^2\sin^3\theta dm}{\Xi^2}+\mathcal{O}(\frac{1}{r^2}) \, ,
\eea
and the angular momentum is
\be
J_{NR}=\frac{2\alpha ma\ell^2}{\Xi^2} =M_{NR}a \, .
\ee
The mass $M_{NR}$ and the angular momentum $J_{NR}$ in the neutral case coincide with
the ones presented in \cite{GILovM}.

Next, we consider the conserved charges of the general dyonic rotating black
hole (\ref{CRBHWeylG}). The perturbation of the metric is determined by the
change of the free parameters $(a,m,p,q)$ through
\be
m\rightarrow m+dm \, , \quad a\rightarrow a+da \, ,
\quad p\rightarrow p+dp \, , \quad q\rightarrow q+dq \, ,\nn
\ee
and both the timelike and spacelike Killing vectors are identical with those in the
neutral case. Under these conditions, after a bit computation, we get the energy and
angular momentum
\be
M_{DR}=\frac{\alpha m\ell^2(12+a^2p^2+a^2q^2)}{6\Xi^2} \, , \quad
J_{DR}=\frac{\alpha ma(12\ell^2+p^2+q^2)}{6\Xi^2} \, . \label{MassJofDRWG}
\ee
If the parameters $(\ell, p, q)$ are rescaled as
\be
\ell\rightarrow \frac{1}{\ell} \, ,\quad
p\rightarrow \frac{p}{m} \, , \quad
q\rightarrow \frac{q}{m} \, ,\nn
\ee
the energy $M_{DR}$ and the angular momentum $J_{DR}$ agree with those obtained
via the definition (\ref{LPPCCwyeG}) in \cite{LLuCRotSW}, where it is demonstrated that
both the energy and angular momentum satisfy the first law of thermodynamics.
Such a match arises from that the integral of the quantity $\xi^{[t}\Theta_W^{r]}(g;h)$ and
$\sqrt{-g}K_W^{tr}\big|_{m,a,p,q=0}$ vanish at asymptotical infinity for the dyonic rotating
black hole.

\section{Conserved charges of black holes in Einstein-Gauss-Bonnet gravity}\label{secthree}

In this section, we discuss calculations on the conserved charges of charged
spherically symmetric black holes in $d$-dimensional $(d>4)$ Einstein-Gauss-Bonnet gravity.
The Lagrangian has the form
\bea
\mathcal{L}_{(GB)}&=& \sqrt{-g}(R-2\Lambda +\alpha L_{(GB)}) \, , \nn \\
L_{(GB)}&=& R^{\mu\nu\rho\sigma}R_{\mu\nu\rho\sigma}-4R^{\mu\nu}R_{\mu\nu}
+R^2 \, , \label{LagraGaussB}
\eea
where the Gauss-Bonnet term $L_{(GB)}$ can be interpreted as a quadratic curvature
correction to general relativity and the negative cosmological constant $\Lambda$
is expressed as $\Lambda=-(d-1)(d-2)\ell^2/2$ in terms of the radius $1/\ell$ of
the AdS spacetime. The variation of the Lagrangian (\ref{LagraGaussB})
yields the expression of EOM
\be
\mathcal{E}^{(GB)}_{\mu\nu}=R_{\mu\nu}-\frac{1}{2}g_{\mu\nu}R+\Lambda g_{\mu\nu}
+2\alpha \Big[R_\mu^{~\lambda\rho\sigma}R_{\nu\lambda\rho\sigma}
+RR_{\mu\nu}-2(R^{\rho\sigma}R_{\mu\rho\nu\sigma}+R_{\mu}^{~\lambda}R_{\nu\lambda})
-\frac{1}{4}g_{\mu\nu}L_{(GB)}\Big] \, .
\ee

Like before, we now derive the surface term and the off-shell Noether potentials.
For convenience, we introduce a tensor $P_{(GB)}^{\mu\nu\rho\sigma}$, defined by
\bea
P^{\mu\nu\rho\sigma}&=&
\frac{\partial\mathcal{L}_{(GB)}}{\partial R_{\mu\nu\rho\sigma}} \nn \\
&=&g^{\mu[\rho}g^{\sigma]\nu}+2\alpha\big[
R^{\mu\nu\rho\sigma}-2\big(g^{\mu[\rho}R^{\sigma]\nu}-g^{\nu[\rho}R^{\sigma]\mu}\big)
+Rg^{\mu[\rho}g^{\sigma]\nu}\big] \, . \label{GaussBPtensor}
\eea
It is easy to prove that the tensor $P^{\mu\nu\rho\sigma}$ has the following
properties
\bea
P_{\mu\nu\rho\sigma}&=&-P_{\nu\mu\rho\sigma}
=-P_{\mu\nu\sigma\rho}
=P_{\rho\sigma\mu\nu}\nn \\
P_{\mu[\nu\rho\sigma]}&=&0 \, , \quad
\nabla_\sigma P^{\mu\nu\rho\sigma} =0 \, .
\eea
With the help of the tensor $P^{\mu\nu\rho\sigma}$, the surface term
$\Theta_{(GB)}^{\mu}(g;\delta g)$ and the off-shell Noether
potential $K_{(GB)}^{\mu\nu}$ for Einstein-Gauss-Bonnet gravity are presented as
\be
\Theta_{(GB)}^{\mu}(g;h)= 2P^{\mu\nu\rho\sigma}\nabla_\sigma h_{\nu\rho}
\, , \quad
K_{(GB)}^{\mu\nu}= 2P^{\mu\nu\rho\sigma}\nabla_{[\rho}\xi_{\sigma]}
 \, , \label{GaussBPK}
\ee
respectively, while the ADT Noether $Q_{(GB)}^{\mu\nu}$ is given by
\be
Q_{(GB)}^{\mu\nu}
=\frac{1}{2}\delta K_{(GB)}^{\mu\nu}
+\frac{1}{2}hP^{\mu\nu\rho\sigma}\nabla_{[\rho}\xi_{\sigma]}
-2\xi^{[\mu}P^{\nu]\lambda\rho\sigma}\nabla_\sigma h_{\lambda\rho}
\, , \label{GaussBNoeQ}
\ee
where $\delta K_{(GB)}^{\mu\nu}$, the variation of the off-shell Noether
potential $K_{(GB)}^{\mu\nu}$, is given by Eq. (\ref{VariGaussBK}) in
the appendix.

In the remainder of this section, we utilize the above Noether potentials and
the formulation of the quasi-local
conserved charge (\ref{QdefineAn}) to calculate the energy of the general charged
spherically symmetric black hole in $d$-dimensional $(d>4)$ Einstein-Gauss-Bonnet gravity,
coupled to Maxwell or nonlinear electrodynamics in AdS spacetime. We first consider the
case for the coupling of Maxwell electrodynamics. The metric of the asymptotically AdS
black hole takes the form \cite{GaussBsoBD,GaussBsoWh,GaussBsoWi,GaussBsoCai,GaussBsoCNO}
\bea
ds^2&=&-H(r)dt^2+\frac{dr^2}{H(r)}+r^2d\Omega_{d-2}^2 \, ,
\quad A=\frac{1}{d-3}\frac{q}{r^{d-3}}dt \, , \nn \\
H(r)&=& 1+\frac{r^2[1-U(r)]}{2\tilde{\alpha}} \, , \quad
U(r)=\sqrt{1-4\tilde{\alpha}\ell^2+\frac{4\tilde{\alpha}m}{r^{d-1}}
-2\alpha \frac{d-4}{d-2}\frac{q^2}{r^{2d-4}}} \, ,  \nn \\
d\Omega_{d-2}^2&=&d\theta_{d-2}^2+\sum_{i=1}^{d-3}\sin^2\theta_{i+1}
\cdot\cdot\cdot \sin^2\theta_{d-2}d\theta_i^2 \, , \quad
0\leq\theta_1<2\pi \, , \quad 0\leq\theta_i<\pi
\, , \label{SpBHGaussB}
\eea
where $\tilde{\alpha}=(d-3)(d-4)\alpha$. The total Lagrangian for this black hole solution
is $\mathcal{L}_{(GB)}-\frac{1}{4} \sqrt{-g}F_{\mu\nu}F^{\mu\nu}$.

To get the energy of the black hole (\ref{SpBHGaussB}), we take the same infinitesimal
parametrization of a single parameter path like in Eq. (\ref{InPaSBHWeylG}).
The timelike Killing vector is $\xi_{GB}^\mu=(-1,0,\cdot\cdot\cdot,0)$.
Then a bit calculation gives the $(t,r)$ components of the Noether potentials
$K_{(GB)}^{\mu\nu}$ and $Q_{(GB)}^{\mu\nu}$ as
\bea
\sqrt{-g}K_{(GB)}^{tr}&=&-\frac{\omega_{d-2}}{\tilde{\alpha}(d-4)}r^{d-1}
[-2+(d-2)U(r)]\Big(-1+U(r)+\frac{r}{2}\frac{dU}{dr}\Big)
\, , \nn \\
\sqrt{-g}Q_{(GB)}^{tr}&=&\frac{(d-2)\omega_{d-2}}{8\tilde{\alpha}}r^{d-1}
\Big(\frac{\partial U^2(r)}{\partial m}dm+\frac{\partial U^2(r)}{\partial q}dq\Big)
\, , \label{trNoKQinGaussG}
\eea
where $\omega_{d-2}=\sin\theta_2\sin^2\theta_3\cdot\cdot\cdot\sin^{d-4}\theta_{d-3}
\sin^{d-3}\theta_{d-2}$ is the square root of the determinant of the line element
$d\Omega_{d-2}^2$. Note that $\sqrt{-g}K_{(GB)}^{tr}$ and $\sqrt{-g}Q_{(GB)}^{tr}$
in Eq (\ref{trNoKQinGaussG}) still hold even if $U(r)$ is a more general function
related to $r$, $m$ and $q$. By utilizing Eq. (\ref{QdefineAn}), it is very easy to
obtain the energy
\be
M_{(GB)} =\frac{(d-2)V_{d-2}}{16\pi}m \, , \label{MassSpGaussB}
\ee
where the volume of the $(d-2)$-sphere $V_{d-2}=2\pi^{\frac{d-1}{2}}/ \Gamma(\frac{d-1}{2})$.
The mass (\ref{MassSpGaussB}) coincides with the one via ADT method \cite{DeserT}
or other methods \cite{MasDKOGausB,MasPeGausB,MasOKGausB,MasKOlGausB,MasMNlGausB,
MasCCHOGausB,ConWPang,MasDBSGausB,MasBRGauB}.

From Eq. (\ref{MassSpGaussB}), one sees that the gauge field makes no contribution to the
energy of the static black hole (\ref{SpBHGaussB}). This is attributed to that the term
with the electric charge parameter $q$ in function $U(r)$ falls off much faster than the
one with the mass parameter $m$ when $r\rightarrow\infty$ so that its contribution to the
energy can be neglected. The same situation takes place for the charged spherically symmetric
black hole in $d$-dimensional $(d>4)$ Einstein-Gauss-Bonnet gravity coupled to nonlinear
electrodynamics in \cite{NonlinMO}. The metric of this black hole can be reexpressed as the
same form as Eq. (\ref{SpBHGaussB}) except for the term associated with the electric field
in $U(r)$, whose contribution to $U(r)$ is smaller than the one from the mass term. Therefore,
the term including the mass parameter still plays a dominant role in determining the energy
of the black hole. By utilizing the formulation of the conserved charge
(\ref{QdefineAn}), combined with Eq. (\ref{trNoKQinGaussG}), we obtain the energy of the
static black hole in the case for the nonlinear coupling of electrodynamics, the same
as that in Eq. (\ref{MassSpGaussB}). It also matches the energy in \cite{NonlinMO}.

At the end of this section, it is worth mentioning that the Gauss-Bonnet term $L_{(GB)}$ in
the Lagrangian (\ref{LagraGaussB}) is a surface term in $d=4$ dimensions, namely, it makes
no contribution to the motion equation. This implies that all the black hole solutions in
four-dimensional Einstein gravity are also the ones in Einstein-Gauss-Bonnet
gravity. We have applied the generalized ADT formalism (\ref{QdefineAn}) to compute the masses
and angular momenta of the four-dimensional Kerr(-AdS) and Kerr-Newman black holes corrected
by the Gauss-Bonnet term. However, our results show that this term makes no corrections to all the
conserved charges, compared with their corresponding ones in Einstein gravity.

\section{Conclusions and discussions}

In this paper, we have extended the off-shell generalization of the conventional ADT formalism
proposed in \cite{KimKY} to calculate the quasi-local conserved charges for the most general
charged static and the dyonic rotating black holes with AdS asymptotics in four-dimensional
conformal Weyl gravity, as well as the charged spherically symmetric black holes in higher dimensional
Einstein-Gauss-Bonnet gravity coupled to Maxwell or nonlinear electrodynamics in AdS spacetime.
Our results confirm those through other methods in the literature. To do this, we first
directly vary the Bianchi identity (\ref{BianchI}), together with the help of the property of Killing
vector, to get the off-shell Noether currents (\ref{OffShNouC2}) and (\ref{AnXiNoethC}),
which are formally different from the ones in \cite{KimKY}. But they are actually equal to each
other since both the surface term $\Theta^{\mu}(g;\xi)$ and the Lie derivative of the surface
term $\Theta^\mu(g;\delta g)$ with respect to the Killing vector $\xi$ vanish for the generally
diffeomorphism gravity theory. The Noether current (\ref{OffShNouC2}) makes it natural to
derive the corresponding off-shell Noether potential (\ref{GenOffShNP2}), without a further requirement
of the property for the symplectic current. Next, we present the explicit expressions of the
surface term and Noether potential for Weyl gravity as the ones in Eqs. (\ref{WeylGsurf}) and
(\ref{WeylGPK}). Utilizing these quantities, we obtain the mass (\ref{MasstaticBH}) of the most
general static black hole in four-dimensional Weyl gravity, as well as the mass and angular
momentum in Eq. (\ref{MassJofDRWG}) for the dyonic rotating black hole, although a naive application
of the original ADT method fails to give a finite result in the static case. Finally, as the case
of Weyl gravity, we start by deriving the surface term and the off-shell Noether potential in
Eq. (\ref{GaussBPK}) for the general Lagrangian (\ref{LagraGaussB}) and then utilize them to gain
the energy (\ref{MassSpGaussB}) of the general charged static asymptotically AdS black hole in higher
dimensional Einstein-Gauss-Bonnet gravity coupled to Maxwell or nonlinear electrodynamics. The energy
(\ref{MassSpGaussB}) is independent on the electric parameter due to a fast fall-off of the term
related to electric field.

Weyl gravity and Einstein-Gauss-Bonnet gravity are two typical fourth-order derivative gravity
theories. It has been proposed that the general fourth-order gravity admits a critical theory
in \cite{CritiLP,CritiDLLP}. The application of the ADT formalism to the critical theory
demonstrates that the mass and angular momenta of all asymptotically Kerr-AdS and
Schwarzschild-AdS black holes vanish at the critical point \cite{ConWPang,CritiDLLP}. We expect
to know whether the generalized formulation of the ADT charge supports this or not.

Although the formulation (\ref{QdefineAn}) for the quasi-local conserved charge is defined
by only taking into account of the contribution from the pure gravity part, our analysis on
charged black holes implies that it may be applicable to the black holes with matter fields,
if the terms including matter fields in the given metric fall off fast enough at asymptotic
infinity to ensure that the formulation (\ref{QdefineAn}) are convergent. Otherwise the effect
of matter fields must be considered \cite{CiteHJPY}. For instance, the formulation
(\ref{QdefineAn}) fails to give a finite mass when it is utilized to the charged rotating
G\"{o}del-type black hole \cite{SQWu} in five-dimensional minimal supergravity. What is more,
even if the conserved charge through the expression (\ref{QdefineAn}) is well-defined in the
presence of matter fields, it is possible for one to omit an finite value
\footnote{We have considered the effect from the action
$L_{EM}=\alpha \sqrt{-g}F_{\mu\nu}F^{\mu\nu}$ along the line of \cite{IyerWald,CiteHJPY}.
It makes no contribution to the total conserved charges of the black holes in this work.}
from the actions of
the matter fields. To overcome these, the contribution from the matter fields has to be taken
into account in the future work.

\appendix
\section{Variations of the potentials $K_W^{\mu\nu}$ and $K_{(GB)}^{\mu\nu}$}

In this appendix, we give the explicit expressions for the variation of the
Noether potential $K^{\mu\nu}$ in both the Weyl and Einstein-Gauss-Bonnet gravity theories.
Note that the Riemann curvature in our conventions is defined by $(\nabla_\mu\nabla_\nu
-\nabla_\nu\nabla_\mu)\omega_\rho =R_{\mu\nu\rho}^{~~~~\lambda}\omega_\lambda$,
where $\omega_\rho$ is an arbitrary vector.

For four-dimensional Weyl gravity, $\delta K_W^{\mu\nu}$ reads
\bea
\delta K_W^{\mu\nu}&=&2\alpha \delta C^{\mu\nu\rho\sigma}\nabla_{[\rho}\xi_{\sigma]}
+2\alpha C^{\mu\nu\rho\sigma} (\xi^\lambda\nabla_{[\rho} h_{\sigma]\lambda}
-h_{\lambda[\rho}\nabla_{\sigma]}\xi^\lambda ) \nn \\
&&-4\alpha h_{\sigma\rho} \xi^\rho\Big(\nabla^{[\mu}R^{\nu]\sigma}
+ \frac{1}{6}g^{\sigma[\mu}\nabla^{\nu]}R\Big)
-4\alpha \xi_\sigma \Big(-h^{\rho[\mu}\nabla_\rho R^{\nu]\sigma}
+\nabla^{[\mu}\delta R^{\nu]\sigma}\nn \\
&&+2R^{\rho\sigma} \nabla^{[\mu}h^{\nu]}_{~~\rho}
-R^{\rho[\mu} \nabla^{\nu]}h^{\sigma}_{~\rho}
-R^{\rho[\mu} \nabla_{\rho}h^{\nu]\sigma}
+R^{\rho[\mu} \nabla^{\mid\sigma\mid}h^{\nu]}_{~~\rho}\nn \\
&&-\frac{1}{6}h^{\sigma[\mu}\nabla^{\nu]}R
-\frac{1}{6}g^{\sigma[\mu}h^{\nu]\rho}\nabla_\rho R
+\frac{1}{6}g^{\sigma[\mu}\nabla^{\nu]}\delta R\Big) \, , \label{VariWeylGK}
\eea
where the variation of the Weyl tensor $C^{\mu\nu\rho\sigma}$ is given by
\be
\delta C^{\mu\nu\rho\sigma}=2\big(h^{[\mu\mid[\rho}R^{\sigma]\mid\nu]}
-g^{[\mu\mid[\rho}\delta R^{\sigma]\mid\nu]}\big)
+\frac{1}{3}\big(\delta R g^{\mu[\rho}g^{\sigma]\nu}
-R h^{\mu[\rho}g^{\sigma]\nu}-R g^{\mu[\rho}h^{\sigma]\nu}\big)
+\delta R^{\mu\nu\rho\sigma} \, .
\ee
In the above two equations, $\delta R^{\mu\nu\rho\sigma}$, $\delta R^{\mu\nu}$ and
$\delta R$ are presented by
\bea
\delta R^{\mu\nu\rho\sigma}&=& -h^{\lambda[\mu}R_\lambda^{~\nu]\rho\sigma}
-2h^{\lambda[\rho}R_\lambda^{~\sigma]\mu\nu}
+\nabla^{[\rho}\nabla^{\mid\nu\mid}h^{\sigma]\mu}
-\nabla^{[\rho}\nabla^{\mid\mu\mid}h^{\sigma]\nu} \, ,\\
\delta R^{\mu\nu}&=&-2h^{\rho(\mu}R^{\nu)}_{~~\rho}
+\nabla_{\rho}\nabla^{(\mu}h^{\nu)\rho}
-\frac{1}{2}\big(\nabla^{\rho}\nabla_{\rho}h^{\mu\nu}
+\nabla^\mu\nabla^\nu h \big) \, ,\\
\delta R&=&\nabla_\mu\nabla_\nu h^{\mu\nu}-\nabla^{\rho}\nabla_{\rho}h
-h^{\mu\nu}R_{\mu\nu} \, .
\eea

For Einstein-Gauss-Bonnet gravity, the variation of the off-shell Noether
potential $K_{(GB)}^{\mu\nu}$ takes the form
\bea
\delta K_{(GB)}^{\mu\nu}&=&2\delta P^{\mu\nu\rho\sigma}\nabla_{[\rho}\xi_{\sigma]}
+2P^{\mu\nu\rho\sigma} (\xi^\lambda\nabla_{[\rho} h_{\sigma]\lambda}
-h_{\lambda[\rho}\nabla_{\sigma]}\xi^\lambda ) \nn \\
\delta P^{\mu\nu\rho\sigma}&=&2\alpha\big[\delta R^{\mu\nu\rho\sigma}
+4\big(h^{[\mu\mid[\rho}R^{\sigma]\mid\nu]}
-g^{[\mu\mid[\rho}\delta R^{\sigma]\mid\nu]}\big)
+\delta R g^{\mu[\rho}g^{\sigma]\nu}
-R \big(h^{\mu[\rho}g^{\sigma]\nu}+g^{\mu[\rho}h^{\sigma]\nu}\big)\big]\nn \\
&&-\big(h^{\mu[\rho}g^{\sigma]\nu}+g^{\mu[\rho}h^{\sigma]\nu}\big)
\, . \label{VariGaussBK}
\eea

\section*{Acknowledgments}

JJP would like to thank Professors Rong-Gen Cai, Yu Tian, Shuang-Qing Wu and Xiao-Ning Wu
for valuable discussions. He is also grateful to Institute of Theoretical Physics,
Chinese Academy of Sciences, for their hospitality during the visit this work was done.
This work was supported by the Natural Science Foundation of China under Grant
Nos. 11275157 and 11247225.


\end{document}